\documentclass[pdftex,twocolumn,epjc3]{svjour3}          % twocolumn

\RequirePackage[T1]{fontenc}

\smartqed  % flush right qed marks, e.g. at end of proof

\RequirePackage{graphicx}
\RequirePackage{mathptmx}      % use Times fonts if available on your TeX system
\RequirePackage{flushend}
\RequirePackage[numbers,sort&compress]{natbib}
\RequirePackage[colorlinks,citecolor=blue,urlcolor=blue,linkcolor=blue]{hyperref}

\usepackage{soul,xcolor}
\usepackage[latin9]{inputenc}
\usepackage{amsmath}
\usepackage{amssymb}
\usepackage{amsfonts}
\usepackage{lscape}
\usepackage{hyperref}
\usepackage{url}
\usepackage{color}
\usepackage{bm}
\usepackage{tabularx}
\usepackage{mathtools}
\usepackage{ulem}
\newcommand{\be}{\begin{equation}}
\newcommand{\ee}{\end{equation}}
\newcommand{\ba}{\begin{eqnarray}}
\newcommand{\ea}{\end{eqnarray}}

\renewcommand{\[}{\begin{equation}}
\renewcommand{\]}{\end{equation}}
\def\be{\begin{equation}}
\def\ee{\end{equation}}
\def\bea{\begin{eqnarray}}
\def\eea{\end{eqnarray}}
\def\eqi{\begin{equation}}
\def\eqf{\end{equation}}
\def\eqia{\begin{eqnarray}}
\def\eqfa{\end{eqnarray}}

\definecolor{darkgreen}{rgb}{0,0.6,0}
\definecolor{violet}{rgb}{1.5,0,1.5}

\journalname{Eur. Phys. J. C}

\begin{document}

\setstcolor{red}

\title{Cosmological constraints from angular homogeneity scale measurements}


\author{Xiaoyun Shao\thanksref{e1,addr1}
        \and
        Carlos A. P. Bengaly\thanksref{e2,addr1}
        \and
        Rodrigo S. Gon\c{c}alves\thanksref{e3,addr1,addr2} %etc.
        \and
        Gabriela C. Carvalho\thanksref{e4,addr3}
        \and
        Jailson Alcaniz\thanksref{e5,addr1}
}

%\thankstext[$\star$]{t1}{Thanks to the title}
\thankstext{e1}{e-mail: xiaoyun48@on.br}
\thankstext{e2}{e-mail: carlosbengaly@on.br}
\thankstext{e3}{e-mail: rsousa@on.br}
\thankstext{e4}{e-mail: gabriela.coutinho@fat.uerj.br}
\thankstext{e5}{e-mail: alcaniz@on.br}

\institute{Observat\'orio Nacional, 20921-400, Rio de Janeiro, RJ, Brazil\label{addr1}
          \and
          Departamento de F\'isica, Universidade Federal Rural do Rio de Janeiro, 23897-000, Serop\'edica, RJ,  Brazil\label{addr2}
          \and
          Faculdade de Tecnologia, Universidade do Estado do Rio de Janeiro,  27537-000, Resende, RJ, Brazil\label{addr3}
}

\date{Received: date / Accepted: date}

\maketitle

\begin{abstract}
In this paper, we obtain new measurements of the angular homogeneity scale ($\theta_H$) from the
BOSS DR12 and eBOSS DR16 catalogs of Luminous Red Galaxies of the Sloan Digital Sky Survey. Considering the flat $\Lambda$CDM model, we use the $\theta_H(z)$ data to constrain the matter density parameter ($\Omega_{m0}$) and the Hubble constant ($H_{0}$). We find $H_0 = 65^{+10}_{-7}$ km s$^{-1}$ Mpc$^{-1}$  and $\Omega_{m0}>0.296$. By combining the $\theta_H$ measurements with current Baryon Acoustic Oscillations (BAO) and Type Ia Supernova (SN) data, we obtain $H_{0}= 66.8 \pm 5.0$ km s$^{-1}$ Mpc$^{-1}$ and $\Omega_{m0} = 0.292^{+0.013}_{-0.015}$ ($\theta_H$ + BAO) and $H_{0}=66.8 \pm 5.4 $ km s$^{-1}$ Mpc$^{-1}$ and $\Omega_{m0}=0.331 \pm 0.018$ ($\theta_H$ + SN). We show that $\theta_H$ measurements help break the BAO and SN degeneracies concerning $H_0$, as they do not depend on the sound horizon scale at the drag epoch or the SN absolute magnitude value obtained from the distance ladder method. Hence, despite those constraints are less stringent compared to other probes, $\theta_H$ data may provide an independent cosmological probe of $H_0$ in light of the Hubble tension. For completeness, we also forecast the constraining power of future $\theta_H$ data via Monte Carlo simulations. Considering a relative error of the order of 1$\%$, we obtain competitive constraints on $\Omega_{m0}$ and $H_0$ ($\approx 5\%$ error) from the joint analysis with current SN and BAO measurements.
\end{abstract}

%%%%%%%%%%%%%%%%%%%%%%%%%%%%%%%%%%%%%%%%%%%%%%%%%%%%
\section{Introduction}\label{sec:intro}
%%%%%%%%%%%%%%%%%%%%%%%%%%%%%%%%%%%%%%%%%%%%%%%%%%%%
The $\Lambda$-Cold Dark Matter ($\Lambda$CDM) framework forms the foundation of the Standard Cosmological Model (SCM), describing an expanding universe undergoing a late-time acceleration phase. Despite its success in explaining a wide range of cosmological observations - such as the Cosmic Microwave Background (CMB), the clustering and weak lensing of large-scale structures (LSS), and distances to standard candles like Type Ia Supernovae (SN)~\cite{aghanim2021planck,Brout:2022vxf,eBOSS:2020yzd,DES:2021wwk,ACT:2023kun,Li:2023tui}  - key components of the model, namely Cold Dark Matter (CDM) and dark energy ($\Lambda$), remain mysterious. 

The SCM faces significant theoretical and observational challenges. Theoretical issues include the coincidence and fine-tuning problems~\cite{Weinberg:2000yb}, while observationally, a roughly 5$\sigma$ tension exists between early and late-time measurements of the Hubble constant (see e.g. \cite{DiValentino:2021izs} and references therein). Moreover, recent observations from the Dark Energy Spectroscopic Instrument (DESI) have hinted at the possibility of dynamic dark energy instead of a constant $\Lambda$~\cite{DESI:2024mwx}, though this result remains debated~\cite{DESI:2024aqx,Cortes:2024lgw}. These challenges emphasize the importance of rigorously testing the fundamental assumptions of the SCM as any significant deviations would require a complete reformulation of the cosmological paradigm. 

One such assumption is the Cosmological Principle (CP), which posits that the universe is statistically homogeneous and isotropic at sufficiently large scales \cite{Clarkson:2010uz,Maartens:2011yx,Clarkson:2012bg,Aluri:2023dsf}. One important aspect of the CP concerns the homogeneity scale ($R_H$), a characteristic distance at which the large-scale structure of the universe  -- measured using biased dark matter tracers like galaxies and quasars -- becomes statistically indistinguishable from a random distribution. In such a distribution, homogeneity and isotropy arise naturally, limited only by shot (Poisson) noise. Extensive studies have sought to measure $R_H$, with most of them concluding that it lies within the range of 70-150 Mpc, based on analyses of galaxy and quasar catalogs from large redshift surveys ~\cite{Hogg:2004vw,Sarkar:2009iga,Scrimgeour:2012wt, Pandey:2013xz, Pandey:2015xea, Sarkar:2016fir, Laurent:2016eqo, Ntelis:2017nrj, Goncalves:2018sxa, Goncalves:2020erb, Kim:2021osl}. However, some studies have questioned these findings, suggesting that measurements could be biased by the survey window function ~\cite{Labini:2009ke, Labini:2011dv, Park:2016xfp,Heinesen:2020wai}.

An alternative approach involves measuring the angular homogeneity scale ($\theta_H$), which depends solely on the positions of sources in the sky ~\cite{Alonso:2013boa}. Unlike the three-dimensional $R_H$, $\theta_H$ avoids the need for a cosmological model to convert redshifts into distances. Recent surveys have successfully identified and measured $\theta_H$ using various redshift catalogs~\cite{Alonso:2014xca, Goncalves:2017dzs, Andrade:2022imy}.  In particular, it was proposed in~\cite{Ntelis:2018ctq, Ntelis:2019rhj} that $R_H$ could serve as a standard ruler, akin to the sound horizon scale in baryon acoustic oscillation (BAO) measurements. However, subsequent research demonstrated that $R_H$ cannot function as a reliable standard ruler due to its non-monotonic dependence on the matter density parameter~\cite{Nesseris:2019mlr}. In contrast, $\theta_H$ has been shown to be a more robust cosmological probe, as it exhibits no such non-monotonic behavior. Measurements of $\theta_H$ at different redshifts can therefore constrain key cosmological parameters, including the matter density and the Hubble constant, as demonstrated in ~\cite{Shao:2023sxk}.

Building on the analysis reported in~\cite{Shao:2023sxk}, this paper introduces a more refined method for measuring $\theta_H$ and extends the cosmological constraints
analysis carried out previously. Moreover, this study derives a larger sample of $\theta_H$ measurements from the SDSS-III BOSS DR12 CMASS galaxy sample of Luminous Red Galaxies (LRGs) and the SDSS-IV eBOSS DR16 LRG catalog, spanning a redshift range of $0.46 \leq z \leq 0.74$. The analysis also combines these measurements with other cosmological datasets, including SN and BAO, to explore their combined constraints. Furthermore, we also evaluate the potential of future $\theta_H$ measurements, expected from ongoing and upcoming redshift surveys, to impose tighter constraints on cosmological parameters.

This paper is organized as follows: In Section \ref{sec:Dataset} we describe the observational data utilized in our analysis. In Section \ref{sec:method}, we present our methodology and our set of angular homogeneity scale measurements. In Section \ref{sec:result}, we show the results of our cosmological analysis, and we present our conclusions and final remarks in Section \ref{sec:conclu}.

%%%%%%%%%%%%%%%%%%%%%%%%%%%%%%%%%%%%%%%%%%%%%%%%%%%
\section{Data}\label{sec:Dataset}
%%%%%%%%%%%%%%%%%%%%%%%%%%%%%%%%%%%%%%%%%%%%%%%%%%%

The Sloan Digital Sky Survey (SDSS) is a global scientific collaboration that has produced highly accurate three-dimensional maps of the Universe. The project was partitioned into four distinct phases: SDSS-I (2000-2005), SDSS-II (2005-2008), SDSS-III (2008-2014), and SDSS-IV (2014-2020). This study focuses on the Data Release 12 (DR12) of the Baryon Oscillation Spectroscopic Survey\footnote{https://live-sdss4org-dr12.pantheonsite.io/} (BOSS)~\cite{dr12a,dr12b,dr12c}, and the Data Release 16 (DR16) of the extended Baryon Oscillation Spectroscopic Survey\footnote{https://www.sdss4.org/dr16/} (eBOSS)~\cite{dr16}, which are subsets of the Sloan Digital Sky Survey III (SDSS-III) and Sloan Digital Sky Survey IV (SDSS-IV), respectively.

\begin{figure*}[!t]
	\centering
    \includegraphics[width=0.5\textwidth]{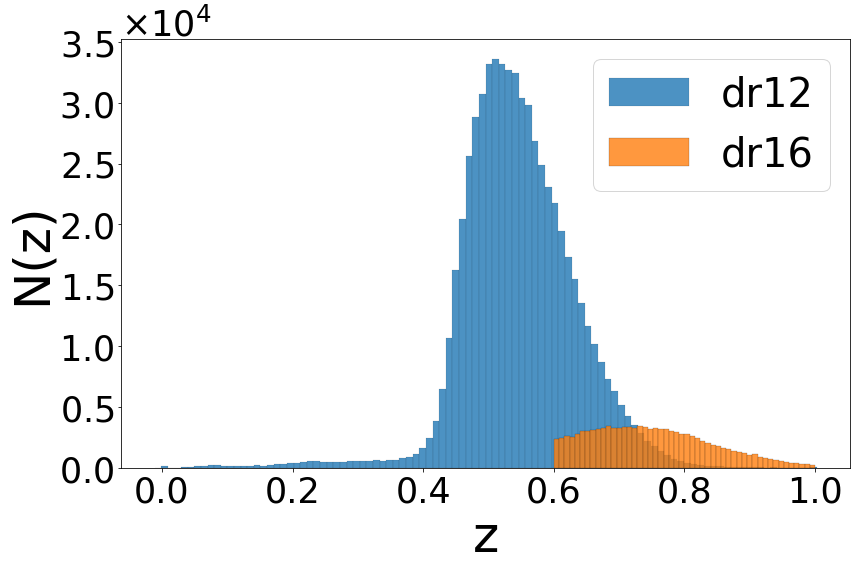} 
    \caption{Redshift distributions of the LRGs from BOSS DR12 and eBOSS DR16 catalogues. We considered the redshift range $0.46 < z < 0.74$.}
\label{fig:hist}
\end{figure*}

\begin{figure*}[!t]
	\centering
    \includegraphics[width=0.45\textwidth]{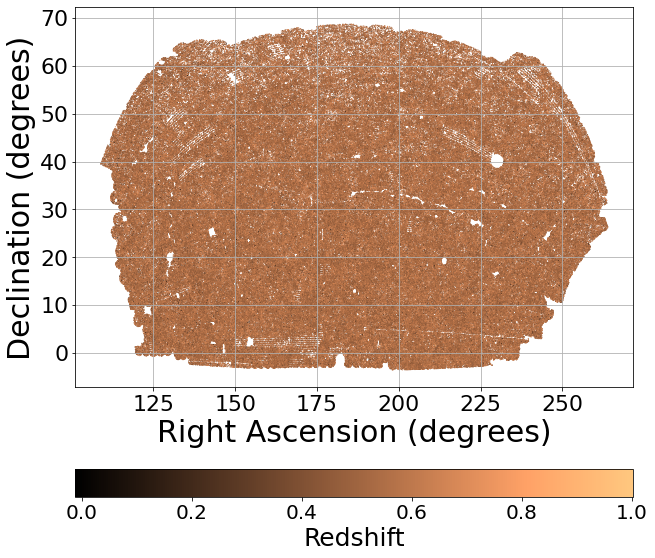} 
    \includegraphics[width=0.45\textwidth]{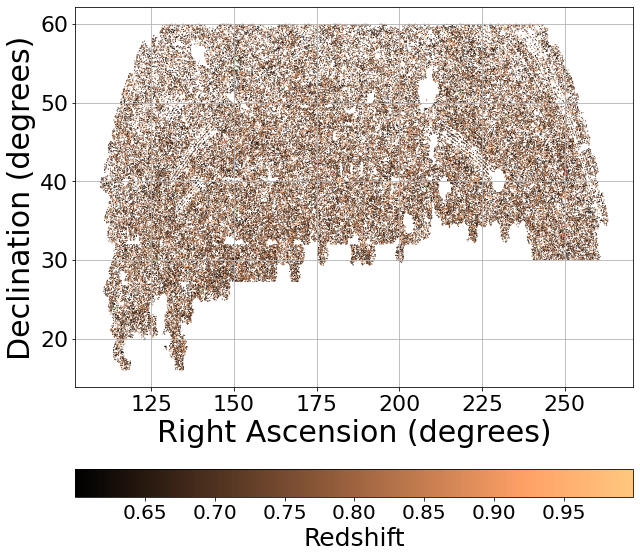} 
    \caption{The footprint of DR12 and DR16, respectively}
\label{fig:dr16sky}
\end{figure*}

Specifically this work adopts the northern sky of both BOSS DR12 and eBOSS DR16 catalogues of Luminous Red Galaxies and for the sake of simplicity we hereafter name them as DR12 and DR16, respectively. The exclusion of the southern sky subsample is due to the limited sky coverage. The main features of the DR12 and DR16 data set used in this work are shown in Table~\ref{t10} and Table~\ref{t20}, respectively. The redshift interval in each dataset is $0.46 < z < 0.62$ with around 420,000 points (DR12) and $0.67 < z < 0.74$ with approximately 30,000 points (DR16), as displayed in Fig.\ref{fig:hist}. The footprint of the sky area coverage of both catalogues is shown in Fig.\ref{fig:dr16sky}.

As we focus on measuring the angular scale of homogeneity, we split the data into redshift bins of $\Delta z = 0.01$. The size of the bin is chosen to avoid projection effects \cite{proj}, so we can avoid possible biases on homogeneity scale measurements, and increase the number of data points, thus providing a good statistical performance for the analysis \cite{Goncalves:2018sxa, Goncalves:2020erb, Alonso:2013boa, Andrade:2022imy}. Also, as shown in~\cite{Shao:2023sxk}, the angular homogeneity scale exhibits a monotonic behavior. for the matter density parameter and Hubble Constant in the range $\Omega_{m0}\in [0.15, 0.8]$ and $H_{0}\in[40, 100]$ (hereafter given in units of km s$^{-1}$ Mpc$^{-1}$, unless stated otherwise), respectively, and it is more sensitive to $H_{0}$ than $\Omega_{m0}$. Thus, this quantity can be used as to constrain those parameters. In order to extend the monotonicity of angular homogeneity scale to lower redshifts, we reduce the range of $\Omega_{m0}$ and $H_{0}$. We find that the angular homogeneity scale presents a monotonic behavior. with $\Omega_{m0}\in [0.15, 0.5]$ and $H_{0}\in[40, 80]$ for $z \gtrsim  0.46$. These intervals of $\Omega_{m0}$ and $H_{0}$ are well within the current uncertainties of their respective constraints obtained by CMB and SN observations, hence we can safely use LRGs within the redshift range $0.46 < z < 0.74$ to perform our analysis.

\begin{table}[!htbp]\centering 
\caption{The redshift bins adopted in our analysis from BOSS (DR12), along with the redshit bin means and corresponding number of LRGs in each bin.}
\label{t10}
%\begin{array}{|c|c|c|}
%\hline 
\begin{tabular}{cccc}
\hline 
z ~~&~~ $\bar{z}$ ~~&~~ $N_{gal}$ ~~\\  \hline ~~ 0.46-0.47 ~~&~~ 0.465 ~~&~~ 22551 ~~\\~~  0.47-0.48 ~~&~~ 0.475 ~~&~~ 27319 ~~\\~~  0.48-0.49 ~~&~~ 0.485 ~~&~~ 29251 ~~\\~~  0.49-0.50 ~~&~~ 0.495 ~~&~~ 31763 ~~\\~~  0.50-0.51 ~~&~~ 0.505 ~~&~~ 33107 ~~\\~~  0.51-0.52 ~~&~~ 0.515 ~~&~~ 32887 ~~\\~~  0.52-0.53 ~~&~~ 0.525 ~~&~~ 32794 ~~\\~~  0.53-0.54 ~~&~~ 0.535 ~~&~~ 31995 ~~\\~~  0.54-0.55 ~~&~~ 0.545 ~~&~~ 31355 ~~\\~~  0.55-0.56 ~~&~~ 0.555 ~~&~~ 29486 ~~\\~~  0.56-0.57 ~~&~~ 0.565 ~~&~~ 28995 ~~\\~~  0.57-0.58 ~~&~~ 0.575 ~~&~~ 25289 ~~\\~~  0.58-0.59 ~~&~~ 0.585 ~~&~~ 23997 ~~\\~~  0.59-0.60 ~~&~~ 0.595 ~~&~~ 22568 ~~\\~~  0.60-0.61 ~~&~~ 0.605 ~~&~~ 20594 ~~\\~~  0.61-0.62 ~~&~~ 0.615 ~~&~~ 18799 ~~\\
%\hline\end{array}
%\end{table}
\hline 
\end{tabular} 
\end{table}

\begin{table}[!htbp]\centering \caption{The redshift bins adopted in our analysis from eBOSS (DR16), along with the redshit bin means and corresponding number of LRGs in each bin.}
\label{t20}
%\begin{array}{|c|c|c|}\hline 
\begin{tabular}{cccc}
\hline 
z ~~&~~ $\bar{ z}$ ~~&~~ $N_{gal}$ ~~\\ \hline ~~ 0.67-0.68 ~~&~~ 0.675 ~~&~~ 4073 ~~\\~~  0.68-0.69 ~~&~~ 0.685 ~~&~~ 4222 ~~\\~~  0.69-0.70 ~~&~~ 0.695 ~~&~~ 4064 ~~\\~~  0.70-0.71 ~~&~~ 0.705 ~~&~~ 4209 ~~\\~~  0.71-0.72 ~~&~~ 0.715 ~~&~~ 4239 ~~\\~~  0.72-0.73 ~~&~~ 0.725 ~~&~~ 4084 ~~\\~~  0.73-0.74 ~~&~~ 0.735 ~~&~~ 4210 ~~\\
%\hline\end{array}
%\end{table}
\hline 
\end{tabular} 
\end{table}

\subsection{ Weights}

{{To correct for known clustering systematics, a specific weight is applied to each galaxy. For the DR12 samples, we followed the weighting scheme described in \cite{reid2016sdss,Ntelis:2017nrj}, where the weight assigned to each galaxy is given by
\begin{equation}
    w_{\text {gal }}=w_{F K P} * w_{\text {systot }} *\left(w_{c p}+w_{n o z}-1\right)
\end{equation}
For the DR16 samples, we adopted the weighting described in~\cite{ross2020completed,bautista2021completed,nadathur2020completed}. Each galaxy is weighted as:
\begin{equation}
w_{\mathrm{tot}}=w_{\mathrm{noz}} w_{\mathrm{cp}} w_{\mathrm{sys}} w_{\mathrm{FKP}}
\end{equation}
where we use the FKP weight, $w_{F K P}$, \cite{feldman1993power} in
order to reduce the variance of the two-point correlation function estimator. $w_{\text {systot }}=w_{\text {star }} * w_{\text {see }}$ is the total angular systematic weight accounting for the seeing effect and the star confusion effect;
$w_{c p}$ accounts for the fact that the survey cannot spectroscopically observe two objects that are closer than $62^{\prime \prime}$ and $w_{noz}$ accounts for redshift failures. }}

%%%%%%%%%%%%%%%%%%%%%%%%%%%%%%%%%%%%%%%%%%%%%%%%%%%%
\section{Methodology}\label{sec:method}
%%%%%%%%%%%%%%%%%%%%%%%%%%%%%%%%%%%%%%%%%%%%%%%%%%%%

\subsection{Theoretical framework}

In this section, we make an overview of the method that was developed to use the angular homogeneity scale $\theta_{H}$ as a cosmological probe. For a complete description, we refer the reader to~\cite{Shao:2023sxk}, where this method is fully described.

The fractal dimension $D_{2}$ is the key quantity needed to measure the angular homogeneity scale of the data, and thus constrain cosmological parameters (see e.g. \cite{Hogg:2004vw,Sarkar:2009iga,Scrimgeour:2012wt, Pandey:2013xz, Pandey:2015xea, Sarkar:2016fir, Laurent:2016eqo, Ntelis:2017nrj, Goncalves:2018sxa, Goncalves:2020erb, Kim:2021osl,Labini:2009ke, Labini:2011dv, Park:2016xfp,Heinesen:2020wai,Alonso:2013boa,Alonso:2014xca, Goncalves:2017dzs, Andrade:2022imy,Ntelis:2018ctq, Ntelis:2019rhj,Nesseris:2019mlr,Shao:2023sxk} and references therein). As we are performing a 2D analysis, the fractal dimension is defined over an angular separation $\theta$. Numerically, we define that the angular homogeneity characteristic scale is reached where $D_{2}$ approaches 1$\%$ of its expected value, following~\cite{Scrimgeour:2012wt}. This implies that $\theta_H$ represents the angular scale where $D_{2}(\theta_H) = 1.98$.

The expression for the fractal dimension is given by
\begin{eqnarray}
\label{eq:d2xi}
D_{2}(\theta)=2+\frac{d \ln}{d \ln \theta}\left[1+\frac{1}{1-\cos \theta} \int_{0}^{\theta} \omega\left(\theta^{\prime}\right) \sin \theta^{\prime} d \theta^{\prime}\right],
\end{eqnarray}
where $\omega(\theta)$ is the two-point angular correlation function (2PACF), which can be expressed as~\cite{crocce2011modelling}
\begin{eqnarray}
\omega(\theta)=\int d z_{1} f\left(z_{1}\right) \int d z_{2} f\left(z_{2}\right) \xi\left(r\left(z_{1}\right), r\left(z_{2}\right), \theta, \bar{z}\right)\label{eq:d2w1}
\end{eqnarray}
where $\xi$ is the three-dimensional two-point correlation function for matter, as described below. Also, we assume $f(z) \equiv b(z) \phi(z)$, where $\phi(z)$ denotes the radial selection, i.e., the probability to include a galaxy in a given redshift bin, and $b(z)$ is the bias parameter, which relates the galaxy distribution to the underlying matter distribution. To account for this bias, a linear bias model designed for the two-point correlation function is applied~\cite{basilakos2008halo}
\begin{equation}
\begin{array}{l}
b(z ; A, B)=A\left(\frac{1+z}{1+z_{\text {eff }}}\right)^{B} \; ,
\end{array}
\label{bz} 
\end{equation}
where $\mathrm{z}_{\text {eff}}$  represents the intermediate redshift bin value. The factor ($1+\mathrm{z}_{\text {eff }}$) is included in order to mitigate the degeneracy that exists between the parameters $A$ and $B$ \cite{ntelis2018scale}. 

In this paper, we consider only top-hat window functions and narrow redshift bins, so we compute the 2PACF and the 3D correlation function for matter as
\begin{equation}
\xi(r)=\frac{1}{2\pi^2}\int_0^\infty j_0(k r) k^2 P(k) dk,\label{eq:corfunc}
\end{equation}
where $P(k)$ is the matter power spectrum and $j_0(x)=\sin{(x)}/x$. Also, the comoving distance to a certain redshift $z$ is given by 
\begin{equation}
r(z)=\int_{0}^{z} \frac{c}{H(z')} dz',
\label{eq:rz}
\end{equation}
where  
\begin{equation}
H(z)=H_0\sqrt{\Omega_{m0}(1+z)^{3} + (1-\Omega_{m0})}\;.
\label{eq:Hz}
\end{equation}

To calculate the theoretical prediction of the non-linear matter power spectrum $P(k)$, we use the {\sc CAMB}~\cite{lewis2000efficient} and HALOFIT codes \cite{smith2003stable} considering the Kaiser effect \cite{kaiser1987clustering}. In the analysis, we disregard the finger-of-God effect \cite{ballinger1996measuring} as it does not affect the calculation of $\theta_H$.

\subsection{Estimating $\theta_H$}

The method above describes how to obtain the theoretical predictions of $\theta_H$. In what follows, we discuss how to estimate the relevant observational quantities to constrain the cosmological parameters.

We assume the Landy-Szalay 2PACF estimator, $\hat{\omega}_{ls}(\theta)$~\cite{landy1993bias}, which is obtained directly from a combination of the data and random catalogues~\cite{Alonso:2013boa,Goncalves:2017dzs,Andrade:2022imy}. It is defined as
\begin{equation}
\label{eq:w_ls}
\widehat{\omega}_{ls}(\theta) = \frac{DD(\theta) -2DR(\theta) + RR(\theta)}{RR(\theta)},
\end{equation}
where $DD(\theta)$, $DR(\theta)$, and $RR(\theta)$ are the numbers of pairs of data as a function of there separations $\theta$, normalized by the total number of pairs, in data-data, data-random and random-random catalogues, respectively. The pair-counting was done with the {\sc TreeCorr} package~\cite{jarvis2004skewness}. Concerning to the random catalogues, we use the BOSS and eBOSS catalogues, which are approximately 50 and 20 times larger than the respective real catalogues~\cite{dr12a,dr12b,dr12c,dr16}. This ensures that the statistical fluctuations caused by the random points are insignificant.

\begin{figure*}[!t]
    \centering
    \includegraphics[width=0.4\textwidth]{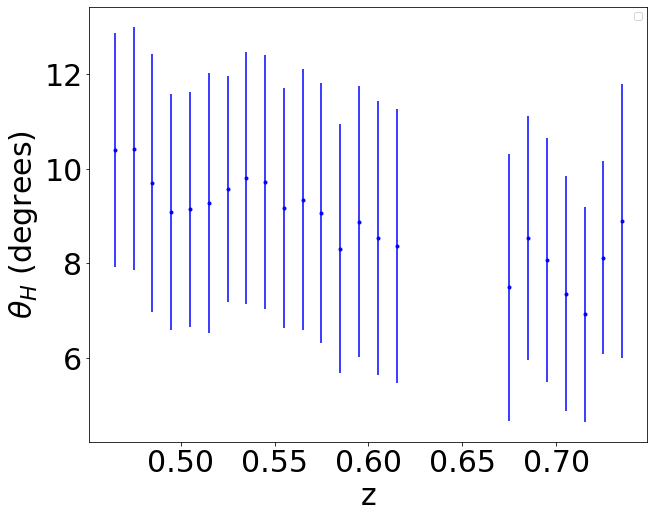} 
    \caption{The evolution of the angular homogeneity scale $\theta_{H}$ with redsfhit. The values of mean and errors bars are taken from Table~\ref{t1} and Table~\ref{t2}.} %(left) and  3D physical homogeneity scale $R_{H}$(right) with redshift are shown.}
\label{fig:thetah_rh_z}
\end{figure*}

Therefore, in order to calculate the $D_2$ values and uncertainties, we perform the following steps: (i) we calculate the value of $D_2$ using Eqs.~\ref{eq:d2xi} and \ref{eq:w_ls} for different values of $\theta$; (ii) we perform a polynomial fit, and determine the angular homogeneity scale $\theta_{H}$ at the point where $D_2=1.98$; (iii) we repeat this procedure 1000 times via bootstrap resampling technique, resulting in a distribution of $\theta_H$ values; (iv) we calculate the mean and standard deviation of the values obtained in step (iii) as our measurements of the angular homogeneity scale and its corresponding uncertainty. The results are displayed in Table~\ref{t1} and Table~\ref{t2} and the respective plot of the redshift evolution of $\theta_{H}$ is presented in Fig.~\ref{fig:thetah_rh_z}.

\begin{table}[t]\centering 
\caption{The redshift means, $\overline{z}$ (each bin has  a width of $\Delta z=0.01$) and the respective observational angular homogeneity scale, $\theta_{H}^{boot}$, obtained from the DR12 data set.}
\label{t1}
\begin{tabular}{cc}
\hline 
$\overline{z}$ ~~&~~ $\theta_{H}(\mathrm{degree})$ ~~\\
\hline
~~ 0.465 ~~&~~ $10.4 \pm 2.47$ ~~\\
~~  0.475 ~~&~~~ $10.43 \pm 2.58$ ~~\\
~~  0.485 ~~&~~ $9.70 \pm 2.73$ ~~\\
~~  0.495 ~~&~~ $9.08 \pm 2.50$ ~~\\
~~  0.505 ~~&~~ $9.14 \pm 2.48$ ~~\\
~~  0.515 ~~&~~ $9.28 \pm 2.76$ ~~\\
~~  0.525 ~~&~~ $9.57 \pm 2.39$ ~~\\
~~  0.535 ~~&~~ $9.80 \pm 2.67$ ~~\\
~~  0.545 ~~&~~ $9.72 \pm 2.69$ ~~\\
~~  0.555 ~~&~~ $9.17 \pm 2.54$ ~~\\
~~  0.565 ~~&~~ $9.35 \pm 2.77$ ~~\\
~~  0.575 ~~&~~ $9.07 \pm 2.75$ ~~\\
~~  0.585 ~~&~~ $8.31 \pm 2.63$ ~~\\
~~  0.595 ~~&~~ $8.88 \pm 2.87$ ~~\\
~~  0.605 ~~&~~ $8.54 \pm 2.90$ ~~\\
~~  0.615 ~~&~~ $8.36 \pm 2.90$ ~~\\ 
\hline 
\end{tabular} 
\end{table}

\begin{table}[t]\centering 
\caption{Same as the previous table, but rather for the DR16 data set.}
\label{t2}
\begin{tabular}{cc}
\hline 
$\overline{z}$ ~~&~~ $\theta_{H}^{ boot}(\mathrm{degree})$ ~~\\
\hline
~~ 0.675 ~~&~~ $7.49 \pm 2.83$ ~~\\
~~  0.685 ~~&~~ $8.53 \pm 2.58$ ~~\\
~~  0.695 ~~&~~ $8.07 \pm 2.59$ ~~\\
~~  0.705 ~~&~~ $7.36 \pm 2.49$ ~~\\
~~  0.715 ~~&~~ $6.92 \pm 2.28$ ~~\\
~~  0.725 ~~&~~ $8.12 \pm 2.04$ ~~\\
~~  0.735 ~~&~~ $8.89 \pm 2.90$ ~~\\ 
\hline 
\end{tabular} 
\end{table}

%%%%%%%%%%%%%%%%%%%%%%%%%%%%%%%%%%%%%%%%%%%%%%%%%%%%
\section{COSMOLOGICAL ANALYSIS}
\label{sec:result}
%%%%%%%%%%%%%%%%%%%%%%%%%%%%%%%%%%%%%%%%%%%%%%%%%%%%

We adopt the Monte Carlo Markov Chain (MCMC) method to constrain cosmological parameters from the angular homogeneity scale measurements $\theta_{H}$. We will now proceed with the next three steps. Initially, we conduct a MCMC analysis to solely the angular scale measurements. Secondly, we combine the $\theta_{H}$ measurements with Supernovae and Baryon Acoustic Oscillations (BAO) data to understand the complementarity of these data and restrict
the cosmological parameters $\Omega_{m0}$ and $H_{0}$. Finally, we derive $\theta_{H}$ points with smaller uncertainties by Monte Carlo Simulations  to forecast the future constraining power of the data.

%------------------------------------------------
\subsection{Angular homogeneity scale  measurements}
\label{data}
%------------------------------------------------

We perform a Markov Chain Monte Carlo analysis to determine the constraints imposed by the angular homogeneity scale measurements on the cosmological parameter space $p_C=(H_{0},\Omega_{m0})$ in the framework of a spatially flat $\Lambda$CDM model. The $\chi^2$ value that we investigate by means of a Markov Chain Monte Carlo (MCMC) approach is defined as\footnote{{We calculate the covariance matrix using the Quick Particle Mesh (QPM) mocks \cite{white2014mock} and the EZmock galaxy catalogues \cite{zhao2021completed}, both of which have been publicly released by the SDSS collaboration. The parameter estimation results based on the covariance matrix show no significant changes. This is because the off-diagonal terms of the covariance matrix are negligible, indicating there is almost no correlation between different redshift bins. This highlights an advantage of using $\theta_{H}$ for cosmological parameter estimation, as it avoid the use of mock catalogs based on a fiducial cosmology, making our method model-independent. Therefore, we can directly use the definition in Eq.~\ref{eq:chi}. }}:
\begin{eqnarray} 
\chi^{2}=\sum_{i=1}^{N}\left(\frac{\mathcal{\theta}_{H}^{G}\left(z_i\right)-\mathcal{\theta}_{H}^{G, t h}\left(z_i ; b, p_{C}, p_{F} \right)}{\sigma_{\mathcal{\theta}_{H}^{G}}(z_i)}\right)^{2}
\label{eq:chi}
\end{eqnarray}
where $\theta_{H}^{G}$ and $\sigma_{\mathcal{\theta}_{H}^{G}}$ correspond to the observational homogeneity scale measurements and their corresponding errors, respectively, as shown in Table~\ref{t1} and Table~\ref{t2}. On the other hand $\theta_{H}^{G,th}$ is the prediction of the homogeneity scale of the galaxies distribution as a function of redshift ($z_i$), the bias factor of the galaxies ($b$, as explicit in eq.~\eqref{bz}) and the set of free $p_{C}$ and fixed $p_{F} $ cosmological parameters. In the present work, we assume $p_{F}= \left(n_{s}, \ln \left[10^{10} A_{s}\right], \Omega_{k}\right)$ as best fit values from Planck 2018~\cite{aghanim2021planck}, so that  the free parameters in our analysis are $\Omega_{m0}$, $H_{0}$, $A$ and $B$.

\begin{figure*}[!t]
\centering
\includegraphics[width=0.46\textwidth]{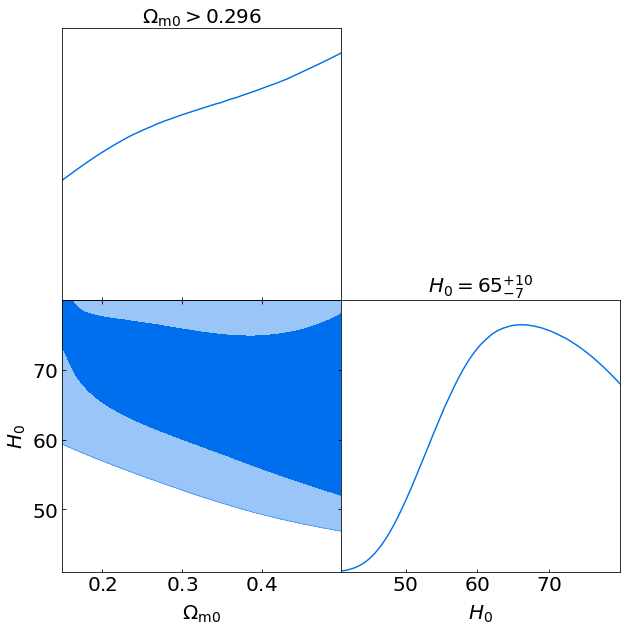}
\includegraphics[width=0.46\textwidth]{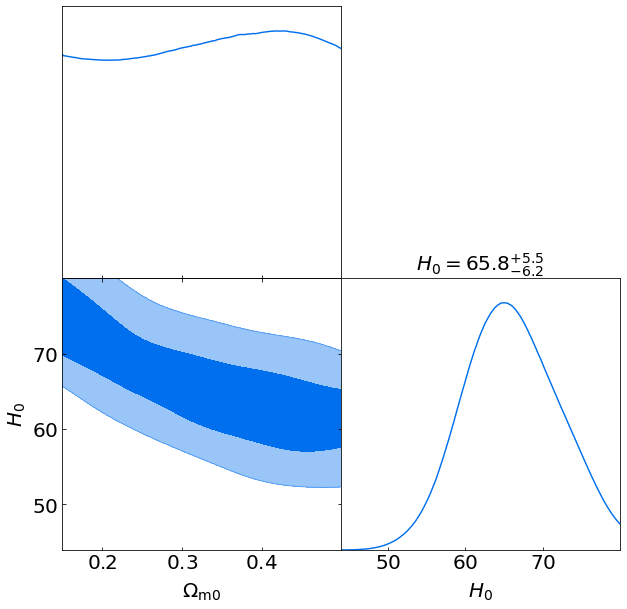}
\caption{The posteriors of $\Omega_{m0}$ and $H_{0}$ with 68$\%$ ($1\sigma$) and 95$\%$ ($2\sigma$) credible regions with the bias parameters $A$ and $B$ (left) being marginalized and (right) fixed at the mean values.}
\label{fig:4p}
\end{figure*}

The priors are flat for all estimated parameters. The
results of the MCMC analysis are presented in Fig.~\ref{fig:4p}. In
the left panel of Fig.~\ref{fig:4p}, we show the constraints imposed
by the $\theta_{H}$ measurements on the cosmological parameters $\Omega_{m0}$ and $H_{0}$ after marginalizing over the bias parameters $A$ and $B$. We estimate $H_{0} = 65^{+10}_{-7}$ $\rm{km.s^{-1}.Mpc^{-1}}$, which agrees with the best-fit from Planck CMB observations, although $H_0$ measurements from SN are well within the $1\sigma$ credible region. It is worth mentioning that given the bigger uncertainty in our estimate compared to Planck CMB and SN, one cannot definitively conclude whether it favors early-time measurements or late-time measurements.

Besides, $\Omega_{m0}$ is mostly unconstrained, {i.e., $\Omega_{m0}>0.296$}, in agreement with the results reported in ~\cite{Shao:2023sxk}. {The correspondent constraints on the bias parameters as free parameters are $A = 2.40 \pm 0.33 $ and $B = 1.15^{+0.80}_{-0.28}$}. \footnote{{{By converting parameters $A$  and $B$ into bias values using Eq.~\ref{bz}, our results are consistent with previous results \cite{Ntelis:2017nrj} within 1$\sigma$ confidence level (CL).}}} We also investigate how these results may change when we fix the bias parameters $A$ and $B$ to the mean values, rather than marginalizing over them. The results are shown in the right panel of Fig.~\ref{fig:4p}. Even though $\Omega_{m0}$ is still unconstrained, we found $H_{0} = 65.8^{+5.5}_{-6.2}$  $\rm{km \cdot s^{-1} \cdot Mpc^{-1}}$, so with a slightly improvement in comparison with the results shown in the left panel of Fig.~\ref{fig:4p}.

{It is important to emphasize that the relative error of $H_0$ obtained from the current data of the angular homogeneity scale need to be improved in order to reach the values of other well stablished cosmological probes, as CMB ($0.7\%$,~\cite{aghanim2021planck}), Cepheids in open clusters ($1.5\%$, \cite{2022CephOpenClust}) or the so-called Tip of the Red Giant Branch (TRGB) ($3.2\%$,~\cite{2021TRG}) but it is still tighter or of the same other than others, as strong-lensing time delay ($9.3\%$,~\cite{2023GL}) and gravitational waves ($13.9\%$,~\cite{2023GW}).}

%%%%%%%%%%%%%%%%%%%%%%%%%%%%%%%%%%%%%%%%%%%%%%%%%%%%%%%%%%%%%%%%%%%%%%%%%%%
% PLOTS

\begin{figure*}[!t]
	\centering
\includegraphics[width=0.46\textwidth]{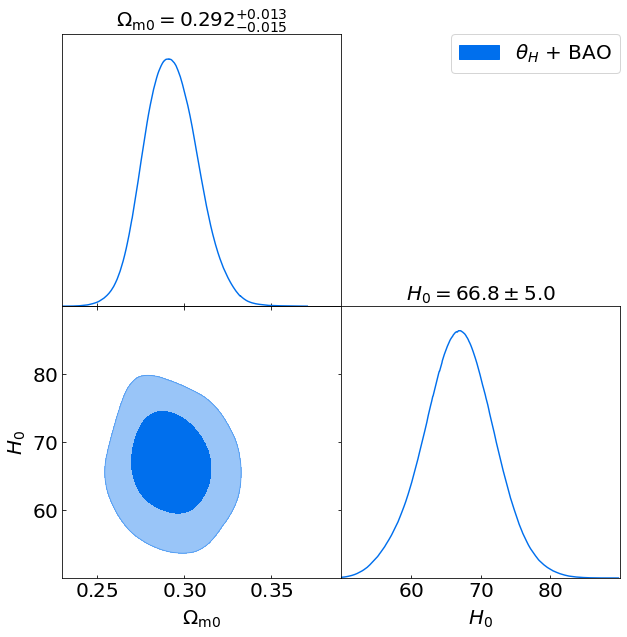}
\includegraphics[width=0.46\textwidth]{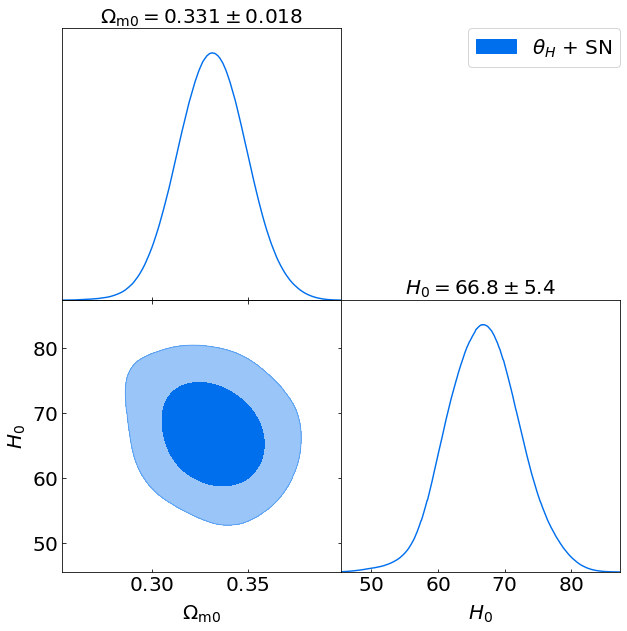}
\caption{Join analysis contours of posteriors for $\Omega_{m0}$ and $H_{0}$, with 68$\%$ and 95$\%$ credible regions, by combining the angular homogeneity scale estimations with BAO (left) and SN (right). In both cases we fix $A$ and $B$ at the mean values.}
\label{fig:comsuphom}
\end{figure*}

\vspace{0.5cm}

%------------------------------------------------
\subsection{Joint analysis}
\label{data_and_others}
%------------------------------------------------

In what follows, we discuss cosmological constraints from joint analyses involving $\theta_{H}$ measurements and the Baryon Acoustic Oscillations (BAO) and Supernova (SN) observations.

As shown earlier, $\theta_{H}$ measurements is more sensitive to $H_{0}$ than $\Omega_{m0}$. In contrast, the opposite occurs for the other two probes considered -- unless we specify priors on $r_{drag}$~\cite{DESI:2024} and $M_B$~\cite{riess2021cosmic,camarena2020local}, representing the sound horizon scale at drag epoch and the SN absolute magnitude, respectively. Hence, a joint analysis between $\theta_{H}$ with BAO or SN may help break the well-known degeneracies between the $H_0$ and $r_{drag}$ in the BAO case and between $H_0$ and $M_B$ in the SN case. This kind of joint analysis could be significant in light of the $\sim 5\sigma$ above tension between early- and late-time measurement of the Hubble Constant. However, we emphasize that our main focus here is to quantify the $\theta_{H}$ performance on constraining the main background parameters of the standard cosmological model, i.e., the $H_{0}$ than $\Omega_{m0}$, rather than trying to address this issue using the $\theta_{H}$. Such an analysis will be pursued in future work. It is worth noting that, in this analysis, the bias parameters $A$ and $B$ are fixed at the mean
values previously mentioned. We make this choice to dodge the degeneracy between those parameters and
$\Omega_{m0}$ and optimize the joint analyses.

The BAO dataset corresponds to a compilation of measurements obtained from the SDSS DR7 main galaxy sample \cite{ross2015clustering}, the 6dF galaxy survey \cite{beutler20116df}, as well as the SDSS BOSS DR12 \cite{alam2017clustering} and eBOSS DR16 surveys \cite{alam2021completed}. The result of the joint analysis between $\theta_{H}$ and BAO data is shown in the left panel of Fig.~\ref{fig:comsuphom}, where we obtain the following constraints: $H_{0}= 66.8 \pm 5.0$ $\rm{km.s^{-1}.Mpc^{-1}}$ and $\Omega_{m0} = 0.292^{+0.013}_{-0.015}$. Our results demonstrate a significant improvement compared to those shown in Fig~\ref{fig:4p}, as expected due to the small uncertainties in the BAO measurements. Also, we confirm that this combination of datasets can help break the BAO degeneracy with $H_{0}$ since we did not assume the $r_{drag}$ value here. 

As for our SN sample, we use the Pantheon+ dataset, which comprises 1550 supernovae distributed across the redshift range of $0.001 < z < 2.26$~\citep{brout2022pantheon+}. The results of the combined $\theta_{H}$ and SN samples are shown in the right panel of Fig~\ref{fig:comsuphom}. As we can see, this combination leads to constraints such as $H_{0}=66.8 \pm 5.4$ $\rm{km.s^{-1}.Mpc^{-1}}$ and $\Omega_{m0}=0.331 \pm 0.018$; hence, as in the BAO case, we find that this joint analysis helps break the SN degeneracy with $H_{0}$, since the latter constraints are independent of the SH0ES prior on $M_{B}$.

\begin{figure*}[!t]
\centering
\includegraphics[width=0.44\textwidth]{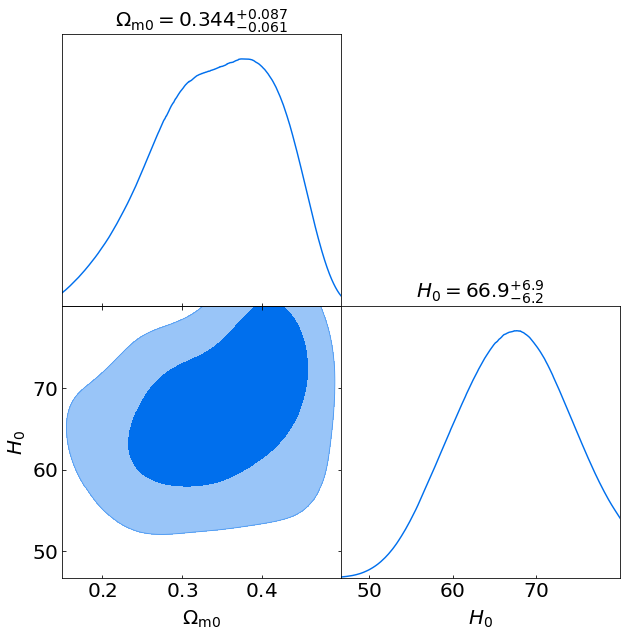}
\caption{The parametric space of $H_0$ and $\Omega_m$ obtained from the simulated data with $1\%$ of current uncertainties, and with bias parameters $A$ and $B$ marginalized over.}
\label{fig:ThetaHSimAlone}
\end{figure*}

\begin{figure*}[!t]
	\centering
\includegraphics[width=0.44\textwidth]{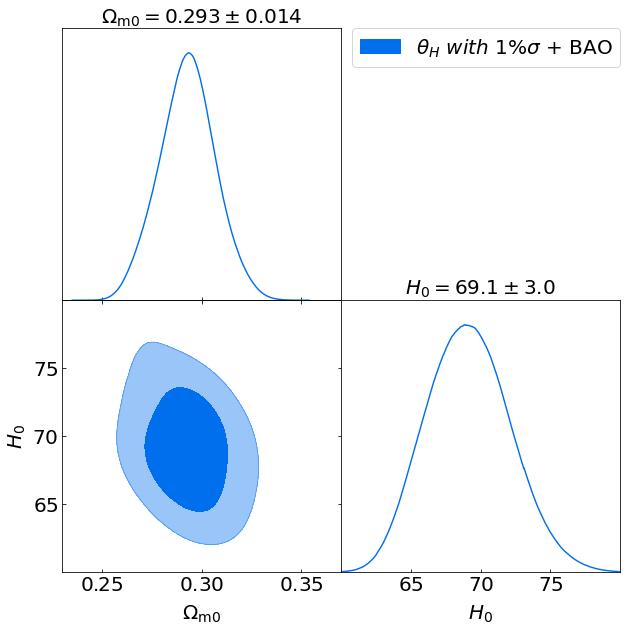}
\includegraphics[width=0.44\textwidth]{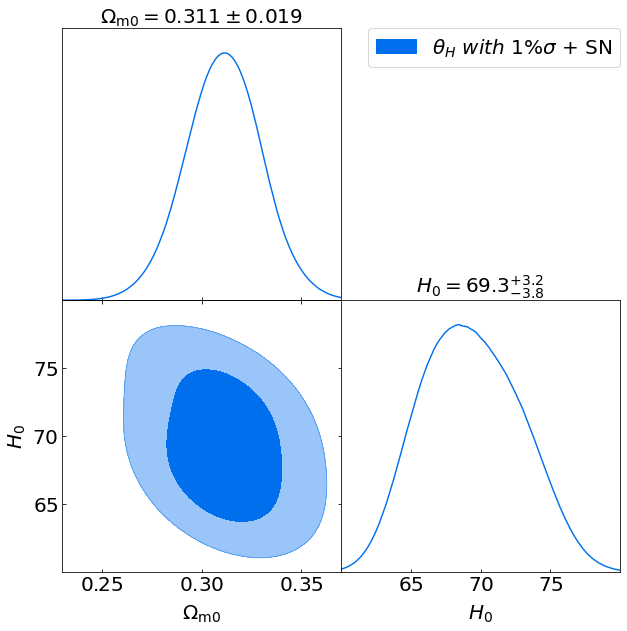}
\caption{Marginalized posteriors on $\Omega_{m0}$ and $H_{0}$ from the combination of the simulated $\theta_{H}$ data assuming $1\%$ precision with BAO (left) and SN (right).}
\label{fig:suphom0p01}
\end{figure*}

%\vspace{0.5cm}

%------------------------------------------------
\subsection{Forecasting $\theta_{H}$}
\label{simu}
%------------------------------------------------

In the previous sections, we showed that the current constraints from $\theta_H$ observations alone is not as competitive as other well-known cosmological probes but can be combined with them to break the degeneracy of some cosmological parameters. In this section, we examine the potential of $\theta_H$ measurements as a robust cosmological probe. {To better understand the effectiveness of future $\theta_H$ data in constraining cosmological parameters, we conduct simulations of the angular homogeneity scale. These simulations assume smaller uncertainties in the measurements and utilize Monte Carlo techniques to enhance accuracy.

Firstly, we fit the observational relative errors $\left( \varepsilon_{\theta_H} = \frac{\sigma_{\theta_H}}{\theta_H} \right)$ from Table~\ref{t1} and Table~\ref{t2}. After that, we assumed a smaller observational uncertainty in order to reproduce expectation values of future measurements. As an example, we choose $\sigma^{sim}_{\theta_H} = 1\% \sigma_{\theta_H} \rightarrow \varepsilon^{sim}_{\theta_H} = 1\% \varepsilon_{\theta_H}$. Afterwards, we then obtain the simulated values for $\theta_{H}$ by sampling them from a Gaussian distribution with a functional form of $N(\theta^{fid}_{H},\sigma^{sim}_{\theta_H})$ at each redshift, where $\theta^{fid}_{H}$ is obtained by assuming a fiducial cosmology model, and the uncertainty $\sigma^{sim}_{\theta_H}$ is obtained as above described. We assume  the flat-$\Lambda$CDM best fit from Planck 2018 as fiducial model and generate data sets with the same features of the observational data, i.e., 23 points and the redshift range from $0.46$ to $0.74$. Then, in order to get the constraints on $H_{0}$ and $\Omega_{m0}$, we perform a MCMC analysis using these simulated data, as in the case of the real data.

From the simulated $\theta_{H}$ data alone, we obtain $H_{0} = 66.9^{+6.9}_{-6.2}$ $\rm{km.s^{-1}.Mpc^{-1}}$ and $\Omega_{m0} = 0.344^{+0.087}_{-0.061}$ (Fig.~\ref{fig:ThetaHSimAlone}). As expected, we find that, as the uncertainty of $\theta_{H}$ decreases, the constraints imposed on $H_{0}$ and $\Omega_{m0}$ are markedly improved -- especially on the $\Omega_{m0}$ parameter, as we obtain non-degenerated constraints of $\Omega_{m0}$ in this case.

We separately perform a joint analysis of those simulated $\theta_{H}$ realizations with the same BAO and SN
external datasets. The results are shown in Fig.~\ref{fig:suphom0p01}. In the left panel of Fig.~\ref{fig:suphom0p01}, we present the constraints obtained from the $\theta_{H}$ simulation combined with the BAO, i.e., $H_{0} = 69.1 \pm 3.0$ $\rm{km.s^{-1}.Mpc^{-1}}$ and $\Omega_{m0} = 0.293 \pm 0.014$. Although the $\Omega_{m0}$ constraint does not change significantly in this case, there is a notable improvement in the $H_{0}$ estimate -- thus helping break the BAO degeneracy between the Hubble Constant and sound horizon scale even further. On the other hand, 
the right panel of Fig.~\ref{fig:suphom0p01} shows the results obtained from the combination of the same simulated $\theta_{H}$ data with the Pantheon+ SN data, resulting in $H_{0}=69.3^{+3.2}_{-3.8}$ $\rm{km.s^{-1}.Mpc^{-1}}$ and $\Omega_{m0}= 0.311 \pm 0.019$. So, combining $\theta_{H}$ with SN leads to tighter constraints on both values of $\Omega_{m0}$ and $H_{0}$, further breaking the SN degeneracy between $M_{B}$ and $H_{0}$.

It is important to remark that, in this joint analysis between $\theta_{H}$ simulations with BAO or SN measurements shown in Fig.~\ref{fig:suphom0p01}, the constraints on $H_0$ and $\Omega_{m0}$ were obtained by marginalizing over the bias parameters, rather than fixing them at the mean values -- as in the Fig.~\ref{fig:comsuphom}. Consequently, the degeneracies related to BAO and SN can be broken even more notably, since we are not making any assumption on the bias model in this case.

%%%%%%%%%%%%%%%%%%%%%%%%%%%%%%%%%%%%%%%%%%%%%%%%%%%
\section{CONCLUSIONS}\label{sec:conclu}
%%%%%%%%%%%%%%%%%%%%%%%%%%%%%%%%%%%%%%%%%%%%%%%%%%%

The large-scale distribution of matter in the Universe has been used as a cosmological probe in many different ways. In this paper, we extended and completed the theoretical and observational studies on the angular homogeneity scale $\theta_H$ reported in previous papers~\cite{Goncalves:2017dzs,Goncalves:2018sxa,Goncalves:2020erb,Andrade:2022imy,Shao:2023sxk}. After deriving new measurements of $\theta_H$, we obtained constraints on the matter density parameter and the Hubble constant from a sample of 23 $\theta_H$ data points lying in the redshift interval $0.465 \leq z \leq 0.735$.

Although the current constraints on $H_0$ and $\Omega_{m0}$ are not competitive with other probes, when combined with BAO and SN observations, the $\theta_H$ data show a complementary behavior with those probes. The results of the joint analysis show a clear improvement in the constraints compared to those obtained with the angular homogeneity scale alone.  This reassures us that $\theta_H$ allows breaking the degeneracies that such probes have concerning $H_{0}$. 

Finally, we also investigated the constraining power of a $\theta_H$ sample with 1\% uncertainty and showed that the limits (mainly) on $\Omega_{m0}$ are significantly improved. With such precision, we also showed that it is possible to improve further the degeneracy breaking on the Hubble constant by combining SN and BAO observations with $\theta_H$ data.

\section*{Acknowledgements}

The authors thank Uendert Andrade for useful discussions. XS is supported by the Coordena\c{c}\~ao de Aperfei\c{c}oamento de Pessoal de N\'ivel Superior (CAPES) - PhD fellowship. CB acknowledges financial support from Funda\c{c}\~ao \`a Pesquisa do Estado do Rio de Janeiro (FAPERJ) - Postdoc Nota 10 (PDR10) fellowship. RSG thanks financial support from FAPERJ grant No. 260003/005977/2024 - APQ1. JSA is supported by CNPq grant No. 307683/2022-2 and FAPERJ grant No. 259610 (2021). This work was developed thanks to the use of the National Observatory Data Center (CPDON).

\bibliographystyle{unsrt}
\bibliography{hom_data_analyze}

%\newpage

%-----------------------------------------%-----------------------------------------

%\onecolumngrid
%\appendix

%-----------------------------------------
%-----------------------------------------

%\newpage

\end{document}